\documentstyle[aps]{revtex}

\def\beq{\begin{equation}}
\def\eeq{\end{equation}}
\def\bea{\begin{eqnarray}}
\def\eea{\end{eqnarray}}
\def\bq{\begin{quote}}
\def\eq{\end{quote}}
\def\PLB{{\it Phys. Lett.} }
\def\PRL{{\it Phys. Rev. Lett.} }

\def\PR{{\it Phys. Rev.} }

\def\bef{\begin{figure}}
\def\befa{\begin{figure}}
\def\enf#1{\label{#1}\end{figure}}
\def\bc{\begin{center}}
\def\ec{\end{center}}
\def\bei{\begin{itemize}}
\def\ei{\end{itemize}} 
\def\be{\begin{eqnarray*}}
\def\ed{\end{eqnarray*}}
\def\bee{\begin{eqnarray}}
\def\eee{\end{eqnarray}}
\def\ben{\begin{equation}}
\def\edn#1{\label{#1}\end{equation}}
\def\bef#1#2{\begin{fmfgraph*}(#1,#2)}
\def\enf{\end{fmfgraph*}}
\def\nn{\nonumber\\}

\def\fr#1#2{\frac{#1}{#2}}

\def\PRL#1#2#3{Phys.Rev.Lett. {\bf #1},#2(#3)}
\def\PRD#1#2#3{Phys.Rev.{\bf D} {\bf #1},#2(#3)}
\def\PR#1#2#3{Phys.Rev.{\bf #1},#2(#3)}
\def\PLB#1#2#3{Phys.Lett. {\bf B #1},#2(#3)}
\def\NPB#1#2#3{Nucl.Phys.{\bf B} {\bf #1},#2(#3)}

\def\g5{\gamma_5}

\def\s#1{\slash\!\!\!{#1}}

\def\->#1{\overrightarrow{#1}}
\def\<-#1{\overleftarrow{#1}}
\def\v0{\->{0}}

\def\aref#1{(\ref{#1})}

\begin{document}
\draft
\title
{A New Formulation for HQET}
\author{Tsung-Wen Yeh}
\address{Institute of Physics, National Chiao-Tung University, Hsinchu, Taiwan 300, R.O.C.}
\maketitle
\begin{abstract}
We proposed a new formulation for heavy quark effective theory (HQET), whose Lagrangian is hermitian and has a manifest reparametrization invariance. As an application, we calculated the semileptonic and nonleptonic inclusive heavy hadron decay rates up to second order mass corrections and found that there are no kinetic energy correction terms.

\end{abstract}

\pagestyle{plain}

\section{Introduction}
The heavy quark effective theory (HQET)\cite{HQET} is an useful tool in studies of heavy quark systems. In infinite heavy quark mass limit, there exists a heavy flavour-spin symmetry of QCD. To derive a HQET theory from QCD, one needs to integrate out the high frequency modes of the heavy quark field to retain the low frequency modes. The high frequency modes with scales larger than two times heavy quark mass can be integrated out by employing the equation of motion (EOM) method \cite{Falk} or the functional integration (FI) method \cite{Mannel}. The Lagrangians of these two HQET theories contain non-hermitian terms which would lead to imaginary mass corrections \cite{Das}. Nevertheless, this problem did not receive too much attentions in literature. We will show in Section II that the non-hermitian terms can be regularized by further integrating out the high frequency modes with scales larger than heavy quark mass.

The heavy quark momentum can be factorized into the heavy quark velocity part and the residual momentum part. Reparameterizing the heavy quark velocity and the residual momentum would not change the effective theory. It implies that the coefficients of mass correction terms in HQET Lagrangian can be fixed by reparameterization \cite{LM}. It was noted that the EOM and FI theories have no manifest invariance under reparameterization \cite{LM,Chen,FGM}. Another HQET theory was derived by employing the Foldy-Wouthuysen (FW) transformation \cite{Balk}. It is unclear that whether the theory is invariant under reparameterization.  

The non-hermitian terms and lacking a manifest reparameterization invariance should be avoided for a self-consistent HQET theory. We will show that a new formulation for HQET could simultaneously resolve these flaws. The remainder of this paper is organized as follows. Section II will explain the cause that leads to the non-hermitian terms, and Section III will present the derivation of a hermitian Lagrangian from QCD. In Section IV, we will show that the hermitian Lagrangian is manifestly invariant under reparameterization. Section V devotes to evaluation of mass corrections for inclusive $B$ decays. An Appendix enumerates the mass correction terms up to $O(1/M_Q^3)$.

\section{The non-hermiticitian terms}
The happens of non-hermitian terms in EOM and FI theories could be better understood by investigating the non-relativistic reduction of the Hamiltonian of an electron interacting with electromagnetic fields. For simplicity, we consider the case that the EM fields are static. The equation of motion for an electron under static Coulomb potentials takes the form
\bee\label{QED1}
i\fr{\partial\psi}{\partial t}=[\vec\alpha\cdot\vec{\pi}+\beta m + e V]\psi\ ,
\eee
where $V$ represents the Coulomb potential, $m$ denotes the electron mass, $\psi$ is the electron wave function, $(\vec\alpha)_i=\gamma^0\gamma^i$ with $i=1,\cdots,3$, $\beta=\gamma^0$ and $\vec{\pi}=-i\vec\nabla$. In the nonrelativistic limit $E\sim m+\vec{p}^2/{2m}$, it is convenient to use the large $\phi$ and small $\chi$ components of $\psi$ defined as
\bee
\psi=(\begin{array}{c}\phi\\\chi\end{array})
\eee
to recast \aref{QED1} into two coupled equations 
\bee
i\fr{\partial\phi}{\partial t}=\vec\sigma\cdot\vec{\pi}\chi+e V\phi+m\phi\ ,\nn
i\fr{\partial\chi}{\partial t}=\vec\sigma\cdot\vec{\pi}\phi+e V\chi-m\chi\ ,
\eee
where we have employed the representation
\bee 
\beta=(\begin{array}{cc}I&0\\0&-I\end{array})\;\;,\;\;
\alpha=(\begin{array}{cc}0&\vec{\sigma}\\\vec\sigma&0\end{array})\ .
\eee
As times evolute, the potential contributions are smeared out by the large value of electron mass, $m$. To avoid this, one may introduce slowly varying functions of times $\Phi$ and $X$ defined as 
\bee
\Phi=e^{imt}\phi\;\;,\;\;X=e^{imt}\chi,
\eee
whose equations are easily derived
\bee\label{int01}
i\fr{\partial\Phi}{\partial t}&=&\vec\sigma\cdot\vec{\pi}X+e V\Phi\ ,\nn
i\fr{\partial X}{\partial t}&=&\vec\sigma\cdot\vec{\pi}\Phi+e VX-2m X .
\eee
Because that $eV\ll 2m$, we are legal to approximate $X$ into 
\bee
X&=&\fr{1}{2m+\pi^0}\vec\sigma\cdot\vec\pi\Phi\nn
&\approx&[\fr{\vec\sigma\cdot\vec\pi}{2m}-\fr{\pi^0\vec\sigma\cdot\vec\pi}{4m^2}+\cdots]\Phi\ ,
\eee
and substitute the expanded $X$ into the first equation of \aref{int01} to obtain
\bee\label{QED3}
i\fr{\partial\Phi}{\partial t}&=&eV\Phi + \fr{(\vec\sigma\cdot\vec\pi)^2}{2m}\Phi
-\fr{e}{4m^2}\{[\vec\sigma\cdot\vec\pi V]\vec\sigma\cdot\vec\pi
+V(\vec\sigma\cdot\vec\pi)^2\}\Phi\ ,
\eee
where 
\bee
[\vec\sigma\cdot\vec\pi V]\vec\sigma\cdot\vec\pi=\vec{\pi}V\cdot \vec{\pi}
+i\vec{\sigma}\cdot(\vec{\pi}V\times \vec{\pi})\ .
\eee
After rewriting \aref{QED3} into Hamiltonian
\bee
H=\Phi^{\dagger}eV\Phi + \Phi^{\dagger}\fr{(\vec\sigma\cdot\vec\pi)^2}{2m}\Phi
-\Phi^{\dagger}\fr{e}{4m^2}\{[\vec\sigma\cdot\vec\pi V]\vec\sigma\cdot\vec\pi
+V(\vec\sigma\cdot\vec\pi)^2\}\Phi\ , 
\eee
we note that the Darwin term in the above Hamiltonian
\bee
O_D=\fr{e}{4m^2}\Phi^{\dagger}\vec{\pi}V\cdot \vec{\pi}\Phi
\eee
is {\bf nonhermitian}
\bee
O_D^{\dagger}=\fr{e}{4m^2}(\vec{\pi}\Phi^{\dagger}\cdot \vec{\pi}V)\Phi\not = O_D\ .
\eee
One can add up $O_D^\dagger$ and $O_D$, and divide the sum by 2 and perform integration by parts by ignoring the surface terms to derive a hermitian Darwin term \cite{Scadron}
\bee
O_D^{R}=\fr{e}{8m^2}\Phi^{\dagger}[(\vec{\pi})^2V]\Phi\ .
\eee
Or equivalently, one can make use of the renormalized $\Phi$ with expression 
\bee
\Phi_{NR}=(1+\fr{(\vec{\pi})^2}{8m^2}+\cdots)\Phi
\eee
to obtain the regularized Darwin term $O_D^{R}$.
The regularized Darwin term $O_D^R$ is in fact the second spatial variations of $V$ due to
the jittery motions of electron with Compton wavelength $\delta{\vec{r}}\sim  1/m$
\bee
<V(\vec{r}+\delta \vec{r})>
\approx <V(r)>+\fr{1}{6m^2}<(\vec{\pi})^2V>\ ,
\eee
where the bracket means integration with electron wavefunctions and the first order variational term vanishes due to the assumption that the wave functions are spherically symmetric.

The above example exhibits the cause of the nonhermitian terms. Similarly, we can show the non-hermitian terms in the HQET theories derived by EOM or FI. The equation of motion for the heavy quark field $\psi$ is
\bee\label{eom1}
(i\s{D}-M_Q)\psi=0\;,
\eee
where $M_Q$ denotes the heavy quark mass and $i\s{D}$ is the covariant derivative $i\s{D}=i\s{\partial}-g\s{A}^a T^a$ . At energies much below than $M_Q$ scale, $\psi$ is not an appropriate variable. One invokes field redefinition $Q(x)=\exp{(iM_Q v\cdot x)}\psi(x)$ to remove the large phase factor $M_Q v$ from the wave function. The $v$ variable represnts the heavy quark velocity. Rewriting (\ref{eom1}) in terms of $Q(x)$ yields
\bee\label{eom1-1}
(i\s{D}-2M_Q\fr{(1-\s{v})}{2})Q=0\ .
\eee
By imposing condition $v^2=1$, one can separate $Q$ into large $h$ and small $H$ components 
\bee\label{eom2}
Q=\fr{1+\s{v}}{2}Q+\fr{1-\s{v}}{2}Q\equiv h + H\;.
\eee 
Substituting \aref{eom2} into \aref{eom1-1} and multiplying $(1-\s{v})/2$ from the left of \aref{eom1-1} yields 
\bee\label{eom3}
H=\fr{1}{2M_Q+i{D}_{\|}}(i\s{D}_{\bot})h
\eee  
with ${D}_{\|}=v\cdot D$ and $\s{D}_{\bot}=\s{D}-\s{v}{D}_{\|}$. Using \aref{eom2} and \aref{eom3} leads to 
\bee\label{eom4}
Q=[1+\fr{1}{2M_Q+i{D}_{\|}}(i\s{D}_{\bot})]h\;.
\eee
Substituting \aref{eom4} into \aref{eom1-1} and expanding it up to $O(1/M_Q^2)$, 
one then arrives at 
\bee\label{eom5}
iD_{\|}h&=&[-\fr{1}{2M_Q}[-D^2_{\perp}+\fr{1}{2}\sigma\cdot G]\nn
&&-\fr{1}{4M_Q^2}[i\sigma_{\alpha\beta}v_{\lambda}G^{\alpha\lambda}D^{\beta}_{\perp}
+iD_{\|}\sigma_{\alpha\beta} G^{\alpha\beta}-iD_{\|}D^2_{\perp}+v_{\alpha}G^{\alpha\beta}D_{\beta}^{\perp}]+O(\fr{1}{M_Q^3})]h\;,
\eee 
where $\gamma^{\mu}\gamma^{\nu}=g^{\mu\nu}+i\sigma^{\mu\nu}$ and $[iD^{\mu},iD^{\nu}]=-iG^{\mu\nu}$ have been used.
Note that the Darwin term (the last term in the second line of \aref{eom5}) is nonhermitian. Following the route in previous QED example, we may regularize the Darwin term by renormalized large components $h^{\prime}$ 
\bee\label{eom6}
h^{\prime}=(1+\fr{1}{8M_Q^2}i\s{D}_{\bot}^2+\dots)h\ .
\eee 
The equation of motion for $h^{\prime}$ thereby takes the form
\bee\label{eom7}
iD_{\|}h^{\prime}&=&[-\fr{1}{2M_Q}[-D_{\perp}^2+\fr{1}{2}\sigma\cdot G]\nn
&&+\fr{1}{8M_Q^2}[i\sigma_{\alpha\beta}v_{\lambda}\{D^{\alpha}_{\perp},G^{\beta\lambda}\}+v_{\alpha}[D_{\beta}^{\perp},G^{\alpha\beta}]
]+O(\fr{1}{M_Q^3})]h^{\prime}\;.
\eee 
The Darwin term in the second line of \aref{eom7} correspond to the relativistic effects of Zitterbeweguen from the jittery motions of the heavy quark with Compton wavelength $\lambda_Q\approx 1/M_Q$. This implies that the large components $h$ still contains high frequency modes whose scales are larger than $M_Q$. These high frequency modes should be integrated out to have a low energy effective theory. It concludes that the mass corrections receive two kinds of contributions: the first kind of contributions comes from the integration out of heavy antiquark degrees of freedom, and the second kind of contributions arises from the integration out of the high frequency modes of the heavy quark degrees of freedom. Only both kinds of contributions having been carefully taking into account can result in a hermitian theory. The integration out of the high frequency modes is equivalent to the renormalization for the heavy quark field \aref{eom6}. This means that the renormalized field $h^{\prime}$ contains only low frequency modes with scales not larger than $M_Q$ and is shwon responsible for low energy physics. A systematic method, which can derive  
a hermitian Lagrangian as well as an appropriate effective field, 
is very intriguing in theory and phenomenology.
To develop this method is the main purpose of this paper.   

To reveal the eligibility of the unrenormalized large components $h$, we support two examples. The first example is the spin sum of $h$ in free theory. By definition \aref{eom4}, the spin sum takes expression
\bee\label{hsum}
\displaystyle{\sum_{\lambda} }h(\lambda)\overline{h}(\lambda)&=&\fr{1+\s{v}}{2}\displaystyle{\sum_{\lambda} }Q(\lambda) \overline{Q}(\lambda)\fr{1+\s{v}}{2}\ ,
\eee
where $\lambda$ denotes the spin of the summed spinors. The spin sum over $Q$ is equal to
\bee\label{Qsum}
\displaystyle{\sum_{\lambda} }Q(\lambda) \overline{Q}(\lambda)=\fr{1+\s{v}}{2}+\fr{\s{k}}{2M_Q}\ ,
\eee
where $k$ means the residual momentum whose magnitude is much smaller than $M_Q$. Substituting \aref{Qsum} into \aref{hsum} yields
\bee
\displaystyle{\sum_{\lambda} }h(\lambda)\overline{h}(\lambda)&=&\fr{1+\s{v}}{2}(1-\fr{\s{k}^2}{4M_Q^2})
\eee
It is noted that the spin sum of the HQET effective spinor $h_v$ is equal to
\bee\label{hvsum}
\displaystyle{\sum_{\lambda} }h_v(\lambda)\overline{h}_v(\lambda)=\fr{1+\s{v}}{2}\;.
\eee
This example shows that the $h$ propagator differs from the $h_v$ propagator
\bee
S_{h_v}=\fr{i}{v\cdot k+i\epsilon}\fr{1+\s{v}}{2}
\eee
by a factor $(1-\fr{\s{k}^2}{4M_Q^2})$, which is just twice the reversal renormalization factor in \aref{eom6}.
We take the free heavy quark spinor as the second example . Let $u_Q$ denote a free full heavy quark spinor whose energy is $E_Q$ and mass $M_Q$ and spatial momentum $\vec{k}$. $u_Q$ can be expressed in terms of its rest frame spinor as
\bee
u_Q=
\left(\begin{array}{cc}
\sqrt{\fr{E_Q+M_Q}{2M_Q}}&\phi^{(\alpha)}\\
\fr{\vec{\sigma}\cdot \vec{k}}{\sqrt{2M_Q(M_Q+E_Q)}}&\phi^{(\alpha)}
\end{array}
\right)
\eee 
where
\bee
\phi^{(1)}=\left(\begin{array}{c}1\\0\end{array}\right)\;\;,\;\;
\phi^{(2)}=\left(\begin{array}{c}0\\1\end{array}\right)
\eee
denote the rest frame spinors.
In the static approximation, we can expand $E_Q$ 
\bee\label{eomc}
E_Q=\sqrt{M_Q^2+\vec{k}^2}\approx M_Q -\fr{\s{k}_{\bot}^2}{2M_Q}\ ,
\eee
where $\vec{k}^2=-\s{k}_{\bot}^2$ and $k_{\bot}=(0,\vec{k})$ have been used. Under this approximation, the full spinor $u_Q$ becomes
\bee\label{eoma}
u_Q=\sqrt{1-\fr{\s{k}_{\bot}^2}{4M_Q^2}}
\left(\begin{array}{c}
\phi^{(\alpha)}\\
\fr{\s{k}_{\bot}}{2M_Q-\s{k}_{\bot}}\phi^{(\alpha)}
\end{array}
\right)\ .
\eee 
From \aref{eoma}, we can identify the large components $h$ and small components $H$ of $u_Q$
\bee\label{eomb}
h&=&\sqrt{1-\fr{\s{k}_{\bot}^2}{4M_Q^2}}\phi\ ,\nn
H&=&\fr{\s{k}_{\bot}}{2M_Q-\s{k}_{\bot}}h\ . \nn
\eee
Spinors $\phi$ and $u_Q$ are well normalized $\phi^{\dagger}\phi=\overline{u}_Qu_Q=1$,
while the large components $h$ has an incorrect normalization 
$\overline{h}h=1-({\s{k}_{\bot}}/{2M_Q})^2$ as pointed out in \cite{Balk}.
Finally, we emphasize that from the equations of motion of $Q$ \aref{eom3} we can directly derive the relation between $h$ and $H$ as  
\bee\label{onshell-1}
H=[\fr{1}{2M_Q-i\s{D}}i\s{D}]h
\eee
and on-shell condition for $Q$
\bee
[iD_{\|} +\fr{(i\s{D})^2}{2M_Q}]Q=0\;.
\eee

\section{Construction of the Effective Theory }
\subsection{Derivation of The Effective Field }
One may match QCD to HQET at the heavy quark mass scale 
by requiring that the 1PI Green's functions of the two
theories describe the same physics.
The simplest way to achieve this is by setting the external quark momenta on shell \cite{FGM}.
The LSZ reduction formula for a heavy quark fermion in momentum space is
\bee
S(P_Q,\dots)&=&\fr{-i}{\sqrt{Z_Q}}\overline{u_Q}(P_Q/M_Q)\fr{(\s{P}_Q-M_Q)}{2M_Q}\dots
\int dx e^{iP_Q\cdot x}<0|T[\psi(x)\dots]|0>|_{P_Q^2=M_Q^2}\nn
&=&\fr{-i}{\sqrt{Z_Q}}\overline{Q}(v+k/M_Q)(\fr{\s{k}}{2M_Q}-\Lambda_v^-)\dots
\int dx e^{ik\cdot x}<0|T[Q(x)\dots]|0>|_{v\cdot k=-k^2/2M_Q}\label{eff0}
\eee
, where $Q(x)=\exp{(iM_Q v\cdot x)\psi(x)}$ and $\psi(x)$ denotes the heavy quark field.
To proceed, we employ the projection operator method  
to derive the matching between the effective spinors $Q$
and $h_v$. $Q$ carries momentum $P_Q=M_Q v +k$,
while $h_v$ has momentum $k$ with respective to a constant moving frame with velocity $v$. Both spinors are equivalent variables for low energy physics.
However, in the limit $M_Q\to \infty$ or at the energy scale much below $M_Q$, $h_v$ is more appropriate than $Q$.  
  
In this method we want to develop, we can specify $Q$ by 
a positive energy projection operator 
\bee
\Lambda^+=\fr{(1+\s{v})}{2}+\fr{\s{k}}{2M_Q}
\eee
which selects spinor $Q$ carrying momentum $M_Q v + k$
\bee
\Lambda^+ Q=Q\label{eff1}\;.
\eee
Eq.(\ref{eff1}) is equivalent to the equation of motion $[i\s{D}-M_Q(1-\s{v})]Q(x)=0$. Being a projection operator, $\Lambda^+$ obeys the identity
\bee
(\Lambda^+)^2=\Lambda^+
\eee 
, which implies on-shell condition
\bee
[{k}_{\|}+\fr{\s{k}^2}{2M_Q}]Q=0\label{eff2}
\eee
with $k_{\|}=v\cdot k$.
The inverse operator of $\Lambda^+$ is the negative energy projection operator 
\bee
\Lambda^-=\fr{(1-\s{v})}{2}-\fr{\s{k}}{2M_Q}
\eee
defined by identity
\bee
 \Lambda^+ +\Lambda^-=1\ .
\eee
In order to derive $h_v$ which respects the physics in the limit $M_Q\to\infty$, we define projectors  
\bee
\Lambda_v^{\pm }=\fr{1\pm\s{v}}{2}\equiv
\displaystyle{\lim_{M_Q\rightarrow\infty}}
\Lambda^{\pm}\ .
\eee
 Note that operators $\Lambda_v^+$ ($\Lambda_v^-$) are the
infinite mass limit of the energy projection operators
$\Lambda^+$($\Lambda^-$).
$\Lambda_v^{\pm }$ satisfy identities
\bee
(\Lambda_v^{\pm })^2=\Lambda_v^{\pm }\ .
\eee
By $1=\Lambda_v^+ + \Lambda_v^-$, we  recast $Q$ to be 
\bee
Q=\Lambda^{+}_{v} Q + \Lambda^{-}_{v} Q = h + H\ .\label{eff2-1}
\eee
From (\ref{eff1}) and (\ref{eff2-1}) we arrive at 
\bee
H=[\fr{1}{2M_Q-\s{k}}\s{k}]h\ ,
\eee
and
\bee
Q=[\fr{1}{1-\fr{\s{k}}{2M_Q}}]h\ .\label{eff3}
\eee
In literature, people always stop at this point and identify $h$ as $h_v$.
However, as point out in last Section, $h$ is not identical to $h_v$.
It is natural to assume that $h_v$ is the limit $M_Q\to\infty$ of $h$
and two spinors are proportional. The first assumption comes from $h_v\equiv\displaystyle{\lim}_{M_Q	\to\infty }Q$
, while the second is based on the fact that
both $h$ and $h_v$ are projected out by $\Lambda_v^+$. 
In this way, we argue that $h=[1+\omega ]h_v$ with 
ansatz $\overline{\omega}=\omega$ and $\s{v}\omega=\omega$.
To derive $\omega$, we note identities 
\bee
\Lambda^+&=&\sum Q \overline{Q}\label{eff4}\ , \\
\Lambda^+_v&=&\sum h_v \overline{h}_v\label{eff5}\ ,
\eee
where summations  over spin indices of the spinors are implied.
Eqs.~\aref{eff4} and \aref{eff5} hold if and only if $\overline{Q}Q=\overline{h}_vh_v=1$.
Eq.~\aref{eff5} comes directly from definition, the limit $M_Q\to\infty$ of \aref{eff4}.
Substituting \aref{eff3} and  $h=[1+\omega]h_v $ into \aref{eff4} and using \aref{eff5} for $h_v$, we get the equation for $\omega$
\bee\label{eff6}
(\omega)^2+2\omega +\fr{1+\s{v}}{2}=(1-\fr{\s{k}}{2M_Q})(\fr{1+\s{v}}{2}+\fr{\s{k}}{2M_Q})
(1-\fr{\s{k}}{2M_Q})\ .
\eee
With the help of on-shell condition \aref{eff2}, \aref{eff6} is recast as
\bee
(\omega)^2+2\omega +(\fr{\s{k}}{2M_Q})^2(\fr{1+\s{v}}{2})=0\ .
\eee
The above equation is easily solved 
\bee
\omega=-1+\sqrt{1+T}
\eee
with $T=-(\fr{\s{k}}{2M_Q})^2(\fr{1+\s{v}}{2})$. We then obtain
\bee
h=\sqrt{1-(\fr{\s{k}}{2M_Q})^2}(\fr{1+\s{v}}{2})h_v\label{eff7}\ .
\eee
The relation between $h$ and $h_v$ is consistent 
with that between $h$ and $h^{\prime}$ found in last Section.  

Combining \aref{eff3} and \aref{eff7}, we get  
\bee
Q=\sqrt{\fr{1+\s{k}/(2M_Q)}{1-\s{k}/(2M_Q)}}\Lambda_v h_v
\equiv\Lambda(w=v+k/M_Q,v)h_v\label{eff8}\;.
\eee
Note that \aref{eff8} is just the Lorentz transformation between two spinors with relative velocity $k/M_Q$. The transformation operator $\Lambda(w=v+k/M_Q,v)$ is identical to the Lorentz boost in the spinor representation \cite{LM}
\bee
\tilde{\Lambda}(w,v)=\fr{1+\s{w}\s{v}}{\sqrt{2(1+v\cdot w)}}\ .\label{eff9}
\eee
In presence of interactions, the Lorentz boot interpretation of \aref{eff8} is invalid.
The reverse transformation from $h_v$ into $Q$ can be derived in a similar way. The result is
\bee
h_v=\sqrt{\fr{1-\s{k}/(2M_Q)}{1+\s{k}/(2M_Q)}}\Lambda^+ Q\ .\label{eff10}
\eee

Transforming \aref{eff8} into coordinate space by replacements $\s{k}\to i\s{D}$, one
obtains 
\bee\label{eff11}
Q(x)=\sqrt{\fr{1+i\s{D}/(2M_Q)}{1-i\s{D}/(2M_Q)}}
\Lambda_v^+h_v(x).
\eee
Field $Q(x)$ is consistent with the field derived by Luke and Manohar
\cite{LM} 
\bee
\Psi_{v}(x)=\tilde{\Lambda}(v+iD/M_Q,v)h_v(x)\label{eff12}\;.
\eee
By employing \aref{eff8} and \aref{eff11}, we arrive at the matching between $Q$ and $h_v$ at
scale equal to $M_Q$ 
\bee\label{match3}
S(k,\dots)
=\fr{-i}{\sqrt{Z_Q}}\overline{h_v}(v)(\fr{\s{k}}{2M_Q}-\fr{1-\s{v}}{2})\dots
\int dx e^{ik\cdot x}\langle 0|T[h_v(x)\dots]|0\rangle|_{v\cdot k=-k^2/2M_Q}\ ,\label{eff9-1}
\eee
which is different from (41) in \cite{KO}
 by a factor $\sqrt{\tilde{Z}(k)}=\sqrt{(1-\s{k}^2/(4M_Q^2)}$. This is because the effective field $h^{KO}_v$ employed in \cite{KO} is the unrenormalized large components $h^{KO}_v=h=\sqrt{\tilde{Z}(k)}h_v$.

By matching QCD and HQET at 2PI and quark-gluon-quark interaction Green functions, we can derive the HQET Lagrangian
\bee\label{eff13}
L&=&\overline{\psi}(i\s{D}-M_Q )\psi\nn 
&=&\overline{Q}(i\s{D}-2M_Q \Lambda_v^- )Q \nn
&=&\overline{h_v}
\Lambda_v^+\sqrt{\fr{1+i\s{D}/(2M_Q)}{1-i\s{D}/(2M_Q )}}
(i\s{D}-2M_Q \Lambda_v^- )\sqrt{\fr{1+i\s{D}/(2M_Q)}{1-i\s{D}/(2M_Q )}}
\Lambda_v^+ h^+_v\ .
\eee
Note that the HQET Lagrangian is hermitian.
This can be inspected by explicitely performing hermitian conjugation
and integration by parts. 

\section{The Velocity Reparameterization transformation}
\subsection{Field Transformation}
The heavy quark momentum $P_Q$ makes no differences for
parameterizations $P_Q=M_Q v +k$ or $P_Q=M_Q v^{\prime}+k^{\prime}$. It was found that the HQET Lagrangian should be invariant under reparameterizations for the velocity $v\to v^{\prime}$ and for the residual momentum $k\to k^{\prime}$. We show the following theorem: If $v$ and $v^{\prime}$ relate each other as 
$v^{\prime}=v+\delta{v}$ with $(v^{\prime})^2=v^2=1$ and $v\cdot\delta{v}+(\delta{v})^2/2=0$,
then $h_{v^{\prime}}$ relates to $h_v$ as
\bee
h_{v^{\prime}}=\sqrt{\fr{1+\delta{\s{v}}/2}{1-\delta{\s{v}}/2}}\Lambda_v^+ h_v\label{RP1}
\eee
and
\bee\label{RP2}
h_{v^{\prime}}(x)=e^{iM_Q \delta{v}\cdot x}\sqrt{\fr{1+\delta{\s{v}}/2}{1-\delta{\s{v}}/2}}\Lambda_v^+ h_v(x)\ .
\eee
Furthermore, if $M_Q\delta{v}=k-k^{\prime}$, 
$Q(P_Q=M_Q v^{\prime}+k^{\prime})$ equals to $Q(P_Q=M_Q v+k)$
and
\bee\label{RP3}
Q(P_Q=M_Q v^{\prime}+k^{\prime})(x)=e^{iM_Q \delta{v}\cdot x}Q(P_Q=M_Q v+k)(x)\ .
\eee
The proof of this theorem is straightforward. Since $v^{\prime}=v+\delta{v}$ and $(v^{\prime})^2=v^2=1$, velocities $v^\prime$ and $v$ have corresponding energy projectors $(1+\s{v}^{\prime})/2$ and $(1+\s{v})/2$, which will project effective fields $h_{v^{\prime}}$ and $h_v$, respectively. 
By replacing $\delta{v}$ with $k/M_Q$ in the transformation \aref{eff8}
, we thus derive the transformation from $h_{v^{\prime}}$ to $h_v$.
The proof of $Q(P_Q=M_Q v^{\prime}+k^{\prime})$ identical to $Q(P_Q=M_Q v+k)$ is also trivial by noting that
\bee
h_{v^{\prime}}&=&\sqrt{\fr{1+\delta{\s{v}}/2}{1-\delta{\s{v}}/2}}
\Lambda_v^+ h_v\ ,\\
Q(P_Q=M_Q v + k)&=&\sqrt{\fr{1+\s{k}/(2M_Q)}{1-\s{k}/(2M_Q)}}
\Lambda_v^+ h_v\ ,\\
Q(P_Q=M_Q v^{\prime}+k^{\prime})&=&\sqrt{\fr{1+\s{k^{\prime}}/(2M_Q)}
{1-\s{k^{\prime}}/(2M_Q)}}\Lambda_{v^{\prime}}^+ h_{v^{\prime}}
\eee
and $M_Q\delta{v}=k-k^{\prime}$. It results in the identity
\bee
Q(P_Q=M_Q v^{\prime}+k^{\prime})&=&\sqrt{\fr{1+\s{k^{\prime}}/(2M_Q)}
{1-\s{k^{\prime}}/(2M_Q)}}\Lambda_{{v}^{\prime}}^+ h_{v^{\prime}}\nonumber\\
&=&\sqrt{\fr{1+\s{k^{\prime}}/(2M_Q)}{1-\s{k^{\prime}}/(2M_Q)}}
\sqrt{\fr{1+\delta{\s{v}}/2}{1-\delta{\s{v}}/2}}\Lambda_v^+ h_v^+\nonumber\\
&=&\sqrt{\fr{1+\s{k}/(2M_Q)}{1-\s{k}/(2M_Q)}}\Lambda_v^+ h_v\nonumber\\
&=&Q(P_Q=M_Q v + k)\ .
\eee
The proof is completed. 
The importance of this transformation is the association of successive transformations.
If we denote the transformation from $h_v$ into $h_{v^{\prime}=v+\delta v}$ by $h_{v^{\prime}}={\it L}(v,v^{\prime})h_v$, then we can have ${\it L}(v,v^{\prime\prime})={\it L}(v^{\prime},v^{\prime\prime}){\it L}(v,v^{\prime})$. We show this explicitely below.
The successive transformations $v\rightarrow v^{\prime}=v+\delta{v_1}$ followed by
 $v^{\prime}\rightarrow v^{\prime\prime}=v^{\prime}+\delta{v_2}=v+\delta{v_1}+\delta{v_2}$
would have the transformations in the effective field as
\bee
h_v\rightarrow h_{v^{\prime}}=
\sqrt{\fr{1+\delta{\s{v}_1}/2}{1-\delta{\s{v}_1}/2}}\fr{1+\s{v}}{2}h_v^+
\eee
and
\bee
h_{v^{\prime}}\rightarrow h_{v^{\prime\prime}}&=&
\sqrt{\fr{1+\delta{\s{v}_2}/2}{1-\delta{\s{v}_2}/2}}\fr{1+\s{v}^{\prime}}{2}h_{v^{\prime}}^+\\
&=&\sqrt{\fr{1+(\delta{\s{v}_1}+\delta{\s{v}_2})/2}
{1-(\delta{\s{v}_1}+\delta{\s{v}_2})/2}}\fr{1+\s{v}}{2}h_{v}^+\ .
\eee

\subsection{Reparameterization Invariance}
We now show that the reparameterization invariance is trivial and manifest for the Lagrangian \aref{eff13}. 
By previous theorem, it is also straightforward to prove that the effective Lagrangian in terms of $Q$ is invariant
under transformations $v\rightarrow v^{\prime}=v+\delta{v}$,  
$M_Q\delta{v}=k-k^{\prime}$ and $Q^+(M_Q v^{\prime}+k^{\prime})(x)=\exp{(iM_Q\delta v\cdot x)}
Q^+(M_Q v+k)(x)$ 
\bee
L^+&=&\overline{Q^+}(P_Q=M_Q v + k)(i\s\!{D}-2M_Q\Lambda_v^-)Q^+(P_Q=M_Q v +k)\nonumber\\
&=&\overline{Q^+}(P_Q=M_Q v^{\prime} + k^{\prime})(i\s\!{D}-2M_Q\Lambda_{v^{\prime}}^-)
Q^+(P_Q=M_Q v^{\prime} +k^{\prime})\;.
\eee
It is also trivial to prove the above invariance for effective Lagrangian in terms of $h_v$ 
\bee
L^+&=&\overline{Q^+}(P_Q=M_Q v^{\prime} + k^{\prime})(i\s\!{D}-2M_Q\Lambda_{v^{\prime}}^-)
Q^+(P_Q=M_Q v^{\prime} +k^{\prime})\nn
&=&\overline{h^+_{v^\prime}}
\Lambda_{v^{\prime}}^+\sqrt{\fr{1+i\s{D}/(2M_Q)}{1-i\s{D}/(2M_Q )}}
(i\s{D}-2M_Q \Lambda_{v^{\prime}}^- )\sqrt{\fr{1+i\s{D}/(2M_Q)}{1-i\s{D}/(2M_Q )}}
\Lambda_{v^{\prime}}^+ h^+_{v^{\prime}}\nn
&=&\overline{h^+_v}
\Lambda_{v}^+
\sqrt{\fr{1+\delta{\s{v}}/2}{1-\delta{\s{v}}/2}}e^{-iM_Q\delta v\cdot x}
\sqrt{\fr{1+i\s{D}/(2M_Q)}{1-i\s{D}/(2M_Q )}}
(i\s{D}-2M_Q \Lambda_{v^{\prime}}^- )\sqrt{\fr{1+i\s{D}/(2M_Q)}{1-i\s{D}/(2M_Q )}}
e^{iM_Q\delta v\cdot x}
\sqrt{\fr{1+\delta{\s{v}}/2}{1-\delta{\s{v}}/2}}
\Lambda_{v}^+ h^+_{v}\nn
&=&\overline{h^+_v}
\Lambda_{v}^+
\sqrt{\fr{1+i\s{D}/(2M_Q)}{1-i\s{D}/(2M_Q )}}
(i\s{D}-2M_Q \Lambda_{v}^- )\sqrt{\fr{1+i\s{D}/(2M_Q)}{1-i\s{D}/(2M_Q )}}
\Lambda_{v}^+ h^+_{v}\nn
&=&\overline{Q^+}(P_Q=M_Q v + k)(i\s\!{D}-2M_Q\Lambda_v^-)Q^+(P_Q=M_Q v +k)
\eee
, where we has used
\bee
 e^{-iM_Q\delta v\cdot x}\sqrt{\fr{1+i\s{D}/(2M_Q)}{1-i\s{D}/(2M_Q )}}
e^{iM_Q\delta v\cdot x}\sqrt{\fr{1+\delta{\s{v}}/2}{1-\delta{\s{v}}/2}}
=\sqrt{\fr{1+i\s{D}/(2M_Q)}{1-i\s{D}/(2M_Q )}}\;.
\eee

\subsection{Zero recoil limit}
The matrix element of $\langle H_v|\bar{Q}\Gamma Q| H_v\rangle$ appears much simplier in the zero recoil limit, $v^\prime=v$. With \aref{eff11} and \aref{match3}, we can derive the following identities
\bee
\langle H_v|\overline{Q}Q|H_v\rangle&=&\langle H_v|\overline{h}_v h_v|H_v\rangle\label{zrl1}\ ,\\
\langle H_v|\overline{Q}\gamma^{\mu}Q|H_v\rangle
&=&\langle H_v|\overline{h}_v (v^{\mu}+\fr{iD^{\mu}}{M_Q})h_v|H_v\rangle\ ,\\
\langle H_v|\overline{Q}\gamma^{\mu}\gamma_5Q|H_v\rangle
&=&\langle H_v|\overline{h}_v(\gamma^{\mu}\gamma_5 -v^{\mu}\fr{i\s{D}}{M_Q}\gamma_5 h_v|H_v\rangle\ ,\\
\langle H_v|\overline{Q}\sigma^{\mu\nu}Q|H_v\rangle
&=&\epsilon^{\mu\nu\alpha\beta}\langle H_v|\overline{h}_v\gamma_{\alpha}\gamma_5(v_{\beta}+\fr{iD_{\beta}}{M_Q}) h_v|H_v\rangle\ .
\eee
Notes that the pseudoscalar currents vanish.

\section{Applications}
We apply the HQET Lagrangian \aref{eff13} to calculate the rates for inclusive semileptonic and nonleptonic heavy hadron decays up to second order mass corrections. 
According to operator product expansion (OPE)\cite{BUV,BSUV,B,MW}, the decay rate of a heavy hadron $H_b$ containing a $b$ quark are expanded in terms of local operators with increasing dimensions   
\bee\label{app1}
\Gamma_{H_b}=\langle H_b|c_1 \bar{b}b
+ c_G \bar{b}\fr{i}{2}\sigma G b+ \cdots|H_b\rangle\ ,
\eee
where $c_1$ and $c_G$ are short distance coefficients. The momentum carried by $H_b$ is chosen as $P_{H_b}=M_{H_b}v$ with $M_{H_b}$ the $H_b$ mass and $v$ its velocity. $|H_b\rangle$ means the eigenstate of the full HQET Lagrangian \aref{eff13} with normalization $\langle H_b(P^{\prime})|H_b(P)\rangle =v^0(2\pi)^3\delta^3(\vec{v}^{\prime}-\vec{v})$. The next step is to expand the matrix elements of the local operators into inverse powers of the $b$-quark mass. Because of \aref{zrl1} the matrix element $\langle H_b|
\bar{b}b|H_b\rangle$ is transformed into $\langle H_b|\bar{h}_v h_v|H_b\rangle$.
There is no mass corrections in $\langle H_b|\bar{b}b|H_b\rangle$. This differs from the conventional result
\bee\label{app2}
\langle H_b|\bar{b}b|H_b\rangle=1-\fr{\mu_\pi^2(H_b)-\mu_{G}^2(H_b)}{2m_b^2}+O(\fr{1}{m_b^3}) \ ,
\eee
where $\mu_\pi^2(H_b)$ and $\mu_{G}^2(H_b)$ parameterize the matrix elements of the kinetic-energy and the chromo-magnetic operators, respectively. The matrix element $\langle H_b|\bar{b}\fr{i}{2}\sigma G b|H_b\rangle$ has coefficient of $O(1/m_b^2)$ and its leading mass corrections are of $O(1/m_b^4)$ will be ignored. With these considerations, we derive the semileptonic decay rate
\bee\label{sl}
\Gamma_{sl}&=&\fr{G_F^2 m_b^5}{192\pi^3}|V_{cb}|^2\left\{
z_0(\fr{m_c}{m_b})- 2z_1(\fr{m_c}{m_b})\fr{\mu_G^2}{m_b^2}+\cdots\right\}\ ,
\eee
which is free from $\mu_\pi$ uncertainty in contrast with the usual formula  
\bee
\Gamma_{sl}&=&\fr{G_F^2 m_b^5}{192\pi^3}|V_{cb}|^2
\left\{z_0(\fr{m_c}{m_b})(1-\fr{\mu_\pi^2-\mu_G^2}{2m_b^2})- 2z_1(\fr{m_c}{m_b})\fr{\mu_G^2}{m_b^2}+\cdots\right\}\ .
\eee
The difference is due to different accounts for the time dilation in the Fermi motion of the heavy quark in the heavy hadron rest frame. Recall that the weak decay widths of muon
\bee
\Gamma(\mu\to e\nu\bar{\nu})=\fr{G_F^2m_\mu^5}{192\pi^3}z_0(\fr{m_e^2}{m_\mu^2})\ .
\eee
One can see that our result is more reasonable because the leading term $\langle H_b|\bar{b}b|H_b\rangle$ is normalized to unity in our HQET theory.
We can derive the nonleptonic decay rate
\bee
\Gamma_{nl}&=&\fr{G_F^2 m_b^5}{192\pi^3}|V_{qb}|^2_{q=c,u}N_c
\left\{w_1 z_0(\fr{m_q}{m_b})- 2w_2(z_1(\fr{m_q}{m_b})+z_2(\fr{m_q}{m_b}) )\fr{\mu_G^2}{m_b^2}\cdots\right\}\ ,
\eee
which is different from the conventional result
\bee
\Gamma_{nl}&=&\fr{G_F^2 m_b^5}{192\pi^3}|V_{qb}|^2_{q=c,u}N_c
\left\{
w_1 z_0(\fr{m_q}{m_b})(1-\fr{\mu_{\pi}^2}{2m_b^2})-2w_2(z_1(\fr{m_q}{m_b})+z_2(\fr{m_q}{m_b}))\fr{\mu_G^2}{m_b^2}\cdots\right\}\ .
\eee
The Wilson coefficients are defined as
\bee
w_1=c_1^2+c_2^2+\fr{c_1c_2}{2N_c}\;\;,\;\;
w_2=\fr{c_1c_2}{2N_c}
\eee
with $c_1=(c_+ +c_2)/2$ and $c_2=(c_+ -c_-)/2$ and the phase space factors are
\bee
z_0(x)=1-8x+8x^3-x^4-12x^2\log{x}\;\;,\;\;
z_1(x)=(1-x)^4\;\;,\;\;
z_2(x)=(1-x)^3\ .
\eee
\section{Conclusion}
We have regularized the non-hermitian terms in the EOM type HQET theories \cite{HQET,Falk,Mannel} up to $O(1/M_Q^2)$. We found that the large components of the heavy quark field are not appropriate variables beyond leading order mass corrections. Only the renormalized large components, whose high frequency modes have been integrated out, can be a relevant effective field for low energies. In terms of the renormalized large components, the HQET Lagrangian \aref{eff13} is hermitian to all orders in $1/M_Q$ and contains manifest reparameterization invariance.

 
\noindent
{\bf Acknowledgments:}
\hskip 0.6cm 
We are grateful for helpful discussions to Guey-Lin Lin and Hsiang-nan Li.
This research was partly supported in part by the National Science Council of R.O.C. under the Grant No. NSC89-2811-M-009-0024.

\appendix
\section{Mass Expansion Lagrangian}
We discuss the mass expansion of the HQET Lagrangian derived in \aref{eff13}.
The HQET Lagrangian $L$ is expanded into mass correction terms as
\bee
L=\displaystyle{\sum_{n=0}^{\infty}}\fr{L^{(n)}}{(2M_Q)^n}
\eee
where the first leading terms $L^{(n)}, n=0,1,2,3$ are enumerated as follows
\bee
L^{(0)}&=&\overline{h}_viD_{\|}h_v\ ,\\
L^{(1)}&=&\overline{h}_v\left [-D_\|^2-D^2+\fr{1}{2}\sigma_{\alpha\beta} G^{\alpha\beta}\right ]h_v\ ,\\
L^{(2)}&=&\overline{h}_v\left [-2iD_{\|}^3+\fr{1}{2}(v_{\alpha}[D_{\beta},G^{\alpha\beta}]+i\sigma_{\alpha\beta}v_{\lambda}\{D^{\beta},G^{\lambda\alpha}\} )\right ]h_v\ ,\\ 
L^{(3)}&=&\overline{h}_v\left [
D^2(D^2+D_{\|}^2)+\fr{1}{2}G^2+\fr{1}{2}\sigma\cdot G D_{\|}^2-\{D^2,\sigma\cdot G \}\right. \nn
&&\hspace{0.5cm}+\sigma_{\alpha\beta}\left (D_{\lambda}\{D^{\beta},G^{\lambda\alpha}\}+[D^{\beta},G^{\lambda\alpha}]D_{\lambda}-iG^{\lambda\alpha}G_{\lambda}^{\ \beta}\right )\nn
&&\left .\hspace{0.5cm}-\fr{i}{4}\gamma_5\epsilon_{\alpha\beta\lambda\rho}G^{\alpha\beta}G^{\lambda\rho}
\right ]h_v\ .
\eee


\begin{thebibliography}{99}
\bibitem{HQET}
N.Isgur and M.B.Wise, \PLB {\bf 232}{113}{1989};{\it idem} {\bf B237}, 527(1990)\\
E.Eichten and B.Hill, \PLB{234}{511}{1990}.\\
B.Grinstein, \NPB{339}{253}{1990}.\\
H.Georgi, \PLB{240}{447}{1990}
\bibitem{Falk}
A.Falk, H.Georgi, B. Grinstein and M. Wise, \NPB{343}{1}{1990}.
\bibitem{Mannel}
T. Mannel, W. Roberts and Z. Ryzak, \NPB{368}{204}{1992}.
\bibitem{Das}
A. Das, {\bf Mod. Phys. Lett. A9},{341}{1994}.
\bibitem{LM}
M.E. Luke and A. V. Manohar, \PLB{286}{348}{1990}.
\bibitem{Chen}
Y.-Q. Chen, \PLB{317}{1993}{421}.
\bibitem{FGM}
Markus Finkemeier, Howard Georgi and Matt McIrvin,\PRD{55}{6933}{1997}.
\bibitem{Balk}
S. Balk, J.G. Korner and D. Pirjol, \NPB{429}{449}{1994}.
\bibitem{Scadron}
Scadron, M. D. Advanced Quantum Theory. New York:Springer-Verlag, 1979. 
\bibitem{KO}
W. Kilian and Ohl, \PLB{279}{194}{1992}.

\bibitem{Sugawara and Okubo}
M. Sugawara and S. Okubo, \PR{117}{605}{1960}.
\bibitem{BUV}
I.I. Bigi, N.G. Uraltsev and A.I. Vainshtein, \PLB{293}{430}{1992}.
\bibitem{BSUV}
I.I. Bigi, M. Shifman, N.G. Uraltsev and A. Vainshtein, \PRL{71}{496}{1993}.
\bibitem{B}
I.I. Bigi et al., in: Proceedings of the Annual Meeting of the Division of Particles and Fields of the APS , Batavia, Illinois,\\ 1992, edited by C. Algright et al.(World Scientific, Singapore, 1993), p. 610.
\bibitem{MW}
A.V. Manohar and M.B. Wise, \PRD{49}{1310}{1994}.
\end{thebibliography}
\end{document}